\documentclass[showkeys,showpacs,superscriptaddress,twocolumn,pra]{revtex4-2}
\usepackage{graphicx}
\usepackage{bm}
\usepackage{color}
\usepackage{inputenc}
\usepackage{subfigure}
\usepackage{amssymb}
\usepackage{amsmath}
\usepackage{ifpdf}
\begin{document}
\title{Fisher information  on the phase transition from regular to flat top optical pulse in cubic-quintic nonlinear media}
\author{Bijan Layek}
\email{bijanlayek1999@gmail.com}
\affiliation{Department of Physics, Kazi Nazrul University, Asansol 713340, W.B., India}
\author{Sk. Siddik}
\affiliation{Department of Physics, Kazi Nazrul University, Asansol 713340, W.B., India}
\author{ Golam Ali Sekh}
\email{golamali.sekh@knu.ac.in} 
 \affiliation{Department of Physics, Kazi Nazrul University, Asansol 713340, W.B., India}
 
\begin{abstract}
\begin{center}
{\bf Abstract}\\
\end{center}
We study the properties of optical pulse  in cubic-quintic nonlinear media using information theoretic approach. The system can support regular (sharp top) and flat top pulse depending on the type of quintic interaction and  the frequency of the pulse.  For defocusing quintic nonlinearity, it holds both regular and flat top soliton. We find that Fisher information suddenly drops near the transition from sharp to flat top pulse. This is also reflected in the linear stability curve where power of the pulse suddenly grows  near the point of transition.  However, the solution is linearly stable according to Vakhitov-Kolokolov criterion. We examine that, in the case of focusing nonlinearity, the change from linearly stable and unstable solitons  becomes imperceptible through Fisher information.
\end{abstract}
\maketitle
\section{Introduction}
If you know less about a system then you can extract more information from it. Thus, more unknown gives more information and more probable provides less information. The challenging task here is the quantification of  information since it needs the information to be a physical quantity\cite{rr1}.  Fortunately, it qualifies the same and thus allows Shannon to introduce a mathematical formulation to measure information. It is the so-called  Shannon information entropy\cite{rr2}. It is  suitable for quantifying global changes in the probability distribution\cite{rr3}. However, the local changes in the characteristics of density distribution $\rho(x)$ can efficiently be identified by Fisher Information (FI) defined by \cite{rr4,rr5}, 
\begin{subequations}
\begin{eqnarray}
F_\rho =\int\frac{1}{\rho(x)}\left[\frac{d\rho(x)}{dx}\right]^2dx
\end{eqnarray}
\begin{eqnarray}
F_\gamma = \int\frac{1}{\gamma(p)}\left[\frac{d\gamma(p)}{dp}\right]^2dp.
\end{eqnarray}
\label{eq1}
\end{subequations}
Here $\gamma(p)$ stands for distribution in momentum space.  In the statistical point of view,  Fisher information can be realized as a measure of disorder or smoothness of the probability density. The FIs in coordinate ($F_\rho$) and momentum $(F_\gamma)$ spaces satisfy an inequality, often called, Cramer-Rao (CR) inequality \cite{rr5,rr6,rr7}. For one-dimensional (1D) system the CR inequality is given by
\begin{equation}
F_\rho F_\gamma\geq 4.
\label{eq3}
\end{equation}
It  serves an uncertainty relation on the measure of Fisher information\cite{rr8}. This inequality gives a lower bound to $F_\rho F_\gamma$.  However, there are some cases  where the above uncertainty relation is not satisfied\cite{rr9,rr10,rr11,rr12,rr13}. Indeed it satisfies the so-called Stam inequality ($F_\rho F_\gamma\leq 4$)\cite{rr14,rr15}. 

Studies based on information theoretic approach   have received a great deal of interest in the last few years\cite{rr16,rr16a}. It can efficiently identify quantum phase in Dicke model\cite{rr17,rr18}, localization of matter waves \cite{rr19}, effects of spin-orbit coupling in ultacold atomic system\cite{rr20}, effects of K-shell electron on neutral atoms\cite{rr21}. 

Objective of the present work is to study properties of regular (RS) and flat top (FTS) soliton and their transition in a purely cubic-quintic nonlinear media  within the framework of information theoretic approach. The FTS appears at certain values of spectral frequencies only in the case of focusing quintic nonlinearity while RS  forms both in the cases of focusing and de-focusing quintic nonlinearities. We have checked that there is a smooth change of FI during the transition from RS to FTS. A sharp change occurs in the pulse power near the transition frequency. However, both RS and FTS are linearly stable according to Vakhitov-Kolokolv criterion\cite{rr22}.

In section II, we present a theoretical model and discuss properties of solutions supported by the medium. More specifically, we discuss how the properties of solutions change for focusing and de-focusing quintic nonlinearity by the variation spectral frequency ($\omega$).  In section III, we calculate Fisher information (FI) and study how FI changes with variation of $\omega$. In section IV, we  examine that the responses of   FI  and linear stability of the pulses are very sensitive near the transition from RS to FTS. In section V, we make some concluding remarks.

\section{Optical pulse in cubic-quintic nonlinear  media}
Let us consider an optical pulse with frequency $\omega$ and intensity $I\propto |E|^2$   is  propagating in a nonlinear medium. For high intensity pulses, higher-order nonlinearities arise due to nonlinear polarization given by
\begin{equation}
P_{NL}=\frac{3}{4}\epsilon_0\chi^{(3)}\vert E\vert^2+\frac{5}{4}\epsilon_0\chi^{(5)}\vert E\vert^4,
\label{eq5}
\end{equation}
where ${\epsilon_0}$ is the permittivity of vacuum and, $\chi^{(3)}$ and $\chi^{(5)}$ are  the third and fifth order susceptibilities which are responsible for cubic and quintic nonlinearities  in fiber.  Thus the  refractive index $n_0(\omega)$ gets modified to $n(\omega)$ :
\begin{equation}
n(\omega,I)=n_0(\omega)+n_2I+n_4I^2.
\label{eq6}
\end{equation}
Using separation of variable one can construct envelop equation of the pulse $E$. In the retarded frame of reference (moving with group velocity $v_g$) the equation of pulse envelope $\psi(x,z)$ can be written as\cite{rr23a}
\begin{equation}
 i\frac{\partial\psi}{\partial z}=\alpha \frac{\partial^2\psi}{\partial x^2}+\kappa\vert\psi\vert^2\psi +\beta\vert\psi\vert^4\psi.
\label{eq5a}
\end{equation}   
Here $\alpha$,  the co-efficient of group velocity dispersion, and $\kappa$ and $\gamma$ represent  strengths of cubic and quintic nonlinearities. Eq.(\ref{eq5a}) is the so-called cubic-quintic nonlinear Schr\"odinger equation (CQNLSE). It is generally used to model  short pulse propagation in optical fiber\cite{rr23}.

We introduce $\psi(z,x)=\phi(x)e^{i\omega z}$ and write 
\begin{equation}
\alpha \frac{d^2 \phi}{d x^2}+\kappa\phi^3+\beta\phi^5=-\omega \phi.
\label{eq6a}
\end{equation}
A particular solution of Eq.(\ref{eq6a}) for $\alpha=-1$, $\kappa=-4$ and $\beta=-3\sigma$ with $\sigma=\pm 1$ is given by \cite{rr24,rr25}
\begin{equation}
\psi(x,z)=\frac{\sqrt{\omega} \,e^{i\omega z}}{{(1+\sqrt{1+\omega\sigma}\cosh(\sqrt{2}\omega x))}^{\frac{1}{2}}}.
\label{eq9}
\end{equation} 
\begin{figure}[h!]
\centering
\includegraphics[scale=.4]{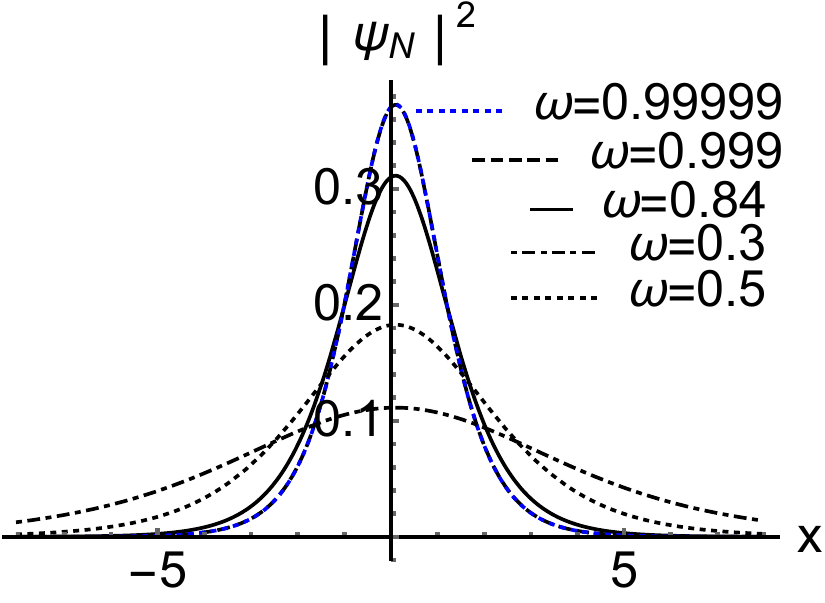}
\vskip 0.2cm
\includegraphics[scale=.4]{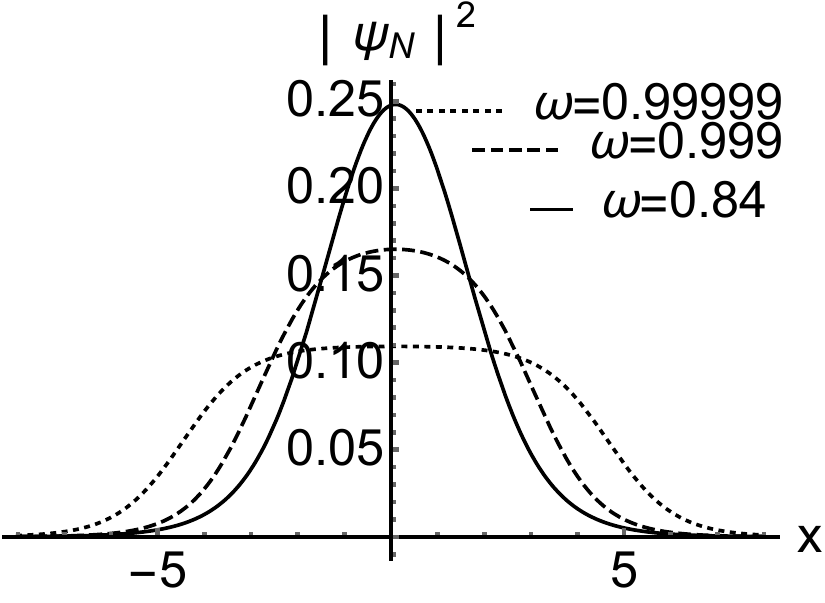}
\caption{Top panel: Density distribution of optical pulse with $\sigma=1$ for different  values of $\omega$. Bottom panel: Density distribution of optical pulse with $\sigma=-1$ for different values of $\omega$.}
\label{fig1}
\end{figure}
We see from Fig.\ref{fig1}  that the density distribution for any value of $\omega$ looks like a regular bright soliton (${sech}$ profile) for $\sigma=+1$ and its peak decreases with the increase of $\omega$ (top panel).  For  $\sigma=-1$ profile peak increases with  $\omega$. However, the peak of the distribution decreases as $\omega$ approaches to $1$ (bottom panel). More specifically, the pulse shape changes from regular soliton (RS) to flat top soliton (FTS). In the next section, we will see how  this change in distribution    reflect in the information measure as well as linear stability of the soliton.

\section{Fisher entropy for focusing and defocusing quintic nonlinearities}
\begin{figure}[h!]
\begin{center}
\includegraphics[scale=.35]{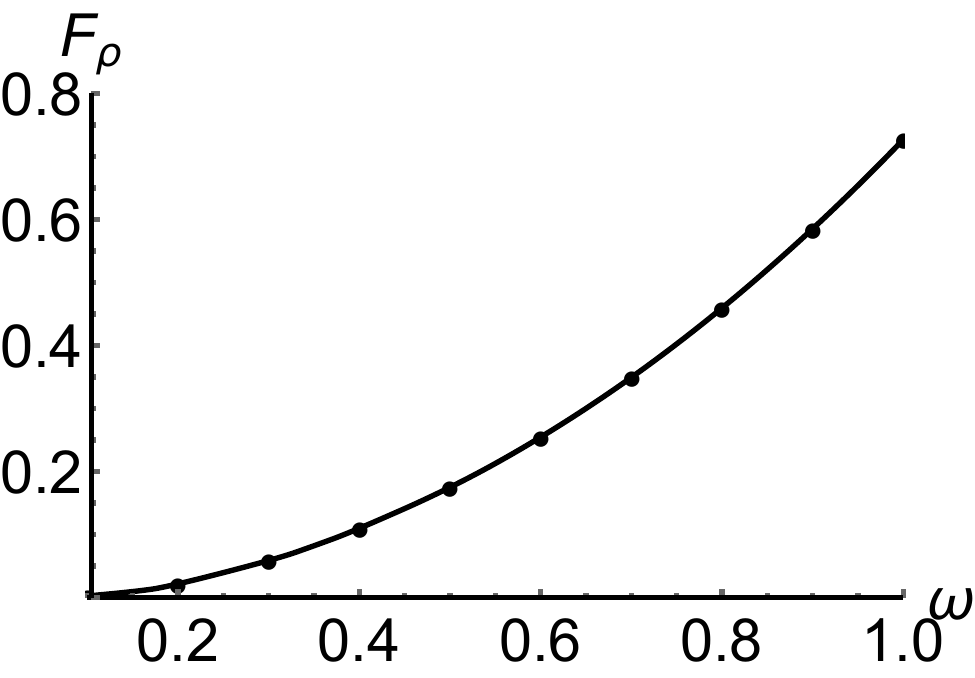}
\caption{Fisher information in position space  for different values of $\omega$ when $\sigma=+1$.}
\label{fig2}
\end{center}
\end{figure}

In order to find Fisher information(FI) and check uncertainty relation, we need to compute  probability density corresponding to  a wavefunction normalized to $1$ both in position and momentum spaces.  But the fact is that the nature of wave function does not allow  straightforward calculation of normalization constant analytically.  In view of this, we make use of numerical integration to compute  Fisher information in position and  momentum spaces. From the numerical results shown in Fig. \ref{fig1}, we see that the distribution sharply changes as $\sigma$ changes from $+1$ to $-1$ in the limit $\omega \rightarrow 1$.  To understand the characteristics of the solution in terms of Fisher information(FI), we first consider the case $\sigma=+1$ and calculate FIs for different values of $\omega$ and display the result in Fig. \ref{fig2}.
We see that the FI increases with the increase of $\omega$ implying the increase of sharpness of the distribution. The changes of FI in momentum space($F_\gamma$) show a trend which is opposite to that in coordinate space (see table I). It is seen that the corresponding entropic uncertainty is always greater than $4$. 
\begin{table}[h]
\centering
\begin{tabular}{|c|c|c|c|}
\hline
 $\omega$ & $F_\rho$ & $ F_\gamma $ & $ F_\rho F_\gamma$ \\
\hline
 0.10 & 0.632477 & 949.371 & 600.455\\
\hline
 0.20 & 0.0218456 & 672.023 & 14.6807\\
\hline
 0.30 & 0.0589059 & 458.846 & 27.0287 \\
\hline
0.40 & 0.110027 & 310.838 & 34.2006 \\
\hline
 0.50 & 0.175046 & 217.369 & 38.0496 \\
\hline
 0.60 & 0.254635 & 157.75 & 40.1687 \\
\hline
 0.70 & 0.349316 & 118.549 & 41.4111  \\
\hline
 0.80 & 0.459421 & 91.827 & 42.1873   \\
\hline
 0.90 & 0.585176 & 72.9771 & 42.7044   \\
\hline
 1.00 & 0.726755 & 59.2551 & 43.0639  \\
\hline
 1.10 & 0.885306 & 48.993 & 43.3738   \\
\hline
 1.20 & 1.05796 & 41.1357 & 43.5199 \\
\hline
\end{tabular}
\caption{Numerical evaluation of Fisher entropy for different value of $\omega$ when $\sigma=+1$.}
\label{table1}
\end{table}

Now we consider the case of  defocusing ($\sigma=-1$) quintic nonlinearity and calculate $F_\rho$ for different values of 
$\omega$.  The result is shown in Fig. 3.  We see that $F_\rho$ suddenly drops as  $\omega$ approaches to $1$. This is an indication of sudden change in properties of the solution. More specifically, the solution changes from regular soliton(RS) to flat top soliton(FTS).  
\begin{figure}[h!]
\begin{center}
\includegraphics[scale=.4]{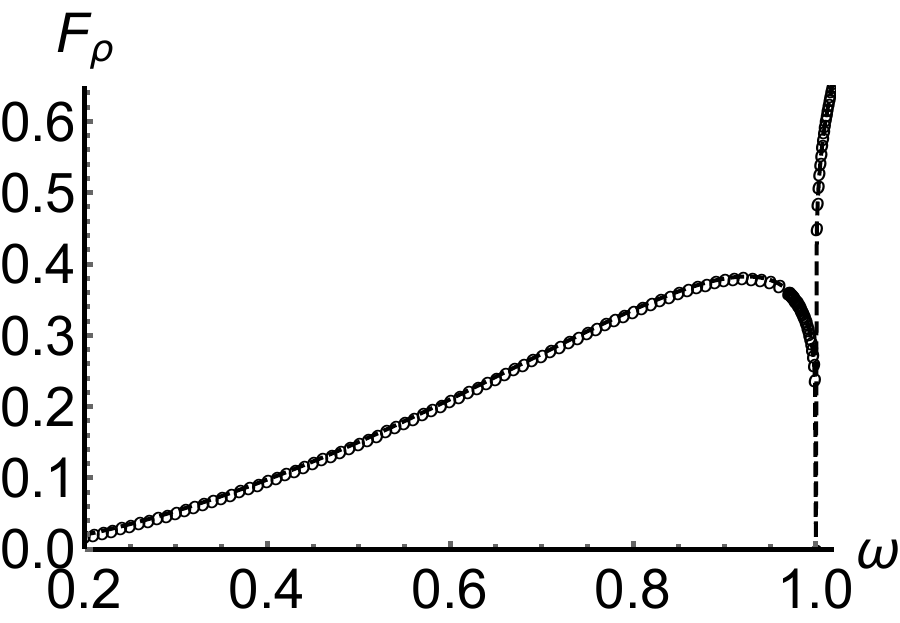}
\vskip 0.1cm
\includegraphics[scale=.4]{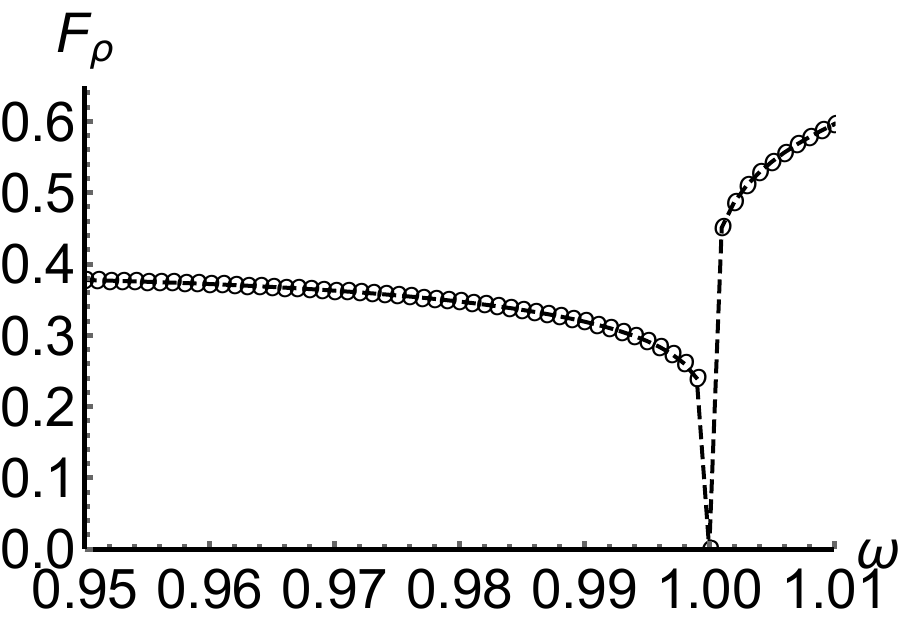}
\caption{Top panel: Fisher entropy in position space  for different values of $\omega$ when $\sigma=-1$. Bottom panel: Close view of $F_\rho$ change near the transition from RS to FTS.}
\label{fig3}
\end{center}
\end{figure}
In Table 2, we display the values of position and momentum spaces Fisher Information and corresponding uncertainty relation. It is clearly seen that the variations of $F_\rho$ and $F_\gamma$ with $\omega$ show opposite trend  and  information uncertainty relation is  satisfied in all the cases.
\begin{table}[ht]
\centering
\begin{tabular}{|c|c|c|c|}
\hline
\ $\omega$ & $F_\rho$ & $ F_\gamma $ & $ F_\rho F_\gamma$ \\
\hline
\ 0.2 & 0.0200466 & 689.996 & 13.8321 \\
\hline
\ 0.3 & 0.053501 & 483.654 & 25.8759 \\
\hline
\ 0.4 & 0.0976709 & 336.604 & 32.8764  \\
\hline
\ 0.5 & 0.150466 & 242.404 & 36.4735  \\
\hline
\ 0.6 & 0.210219 & 181.934 & 38.2459 \\
\hline
\ 0.7 & 0.273954 & 142.582 & 39.0609 \\
\hline
\ 0.8 & 0.33562 & 117.111 &  39.3048 \\
\hline
\ 0.9 & 0.379517 & 103.032 & 39.1024\\
\hline
\ 0.97 & 0.362611 &106.767  & 38.7148 \\
\hline
\end{tabular}
\caption{Numerical evaluation of Fisher Information for different value of $\omega$ when $\sigma=-1$.}
\label{table2}
\end{table}

\section{Vakhitov-Kolokolov criterion and Fisher information}
\begin{figure}[h!]
\begin{center}
\includegraphics[scale=.5]{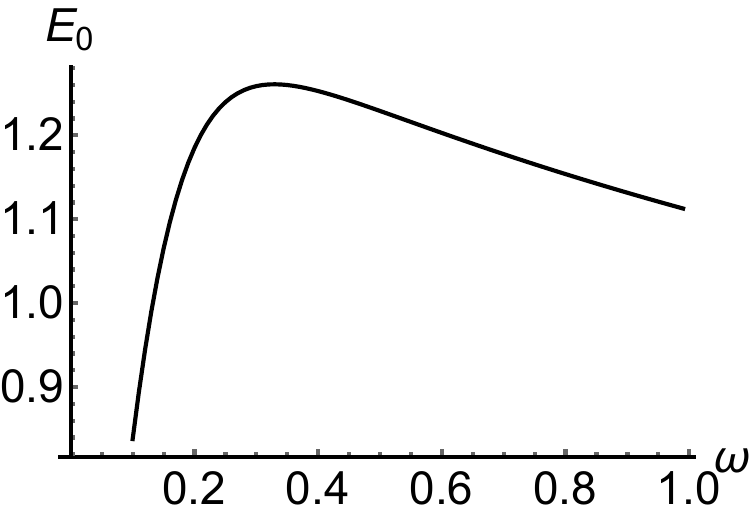}
\vskip 0.1cm
\includegraphics[scale=.5]{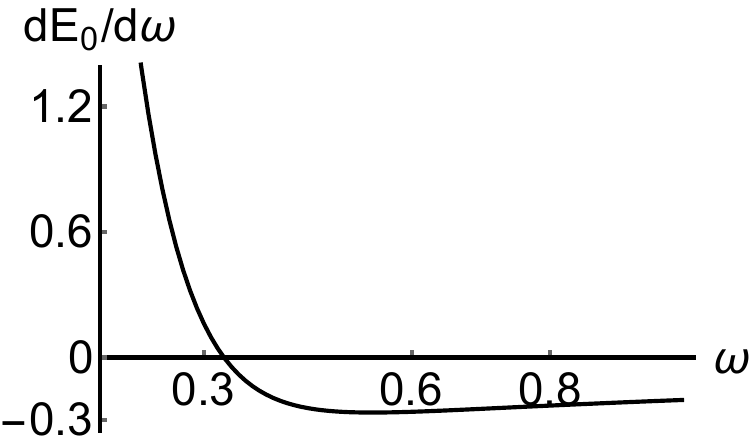}
\caption{Top panel: Variation of $\omega$ with $E_0$ for  $\sigma=1$. Bottom panel:  $dE_0/ d\omega$ versus $\omega$  for $\sigma=1$. }
\label{fig4}
\end{center}
\end{figure}

In this section our main purpose is to check correspondence information theoretic study and linear stability. More specifically, we want see the solutions permitted by entropic uncertainty relation are also linearly stable or not. In view of this, we calculate $E_0(=\int |\psi|^2\,dx)$ and  ${dE_0}/{d\omega}$ numerically for different values of $\omega$.  Fig.\ref{fig4} clearly shows that ${dE_0}/{d\omega}$  changes its sign at $\omega \approx 0.3$ and the corresponding power is $E_0 \approx 1.25$. This implies that the solution tends to become unstable if we increase power and/or of  frequency of the pulse. In both the case, however, $F_\rho F_\gamma \geq 4$. Thus we say that the entropy uncertainly relation can be satisfied even  for unstable solution.

\begin{figure}[h!]
\begin{center}
\includegraphics[scale=.4]{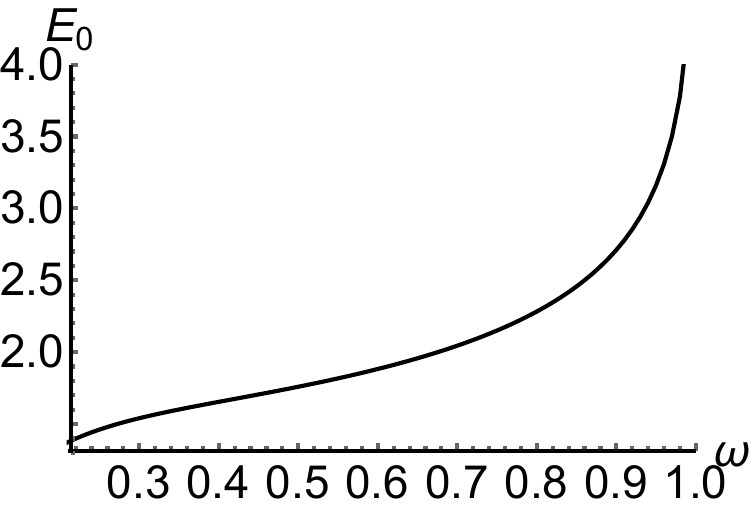}
\vskip 0.175cm
\includegraphics[scale=.4]{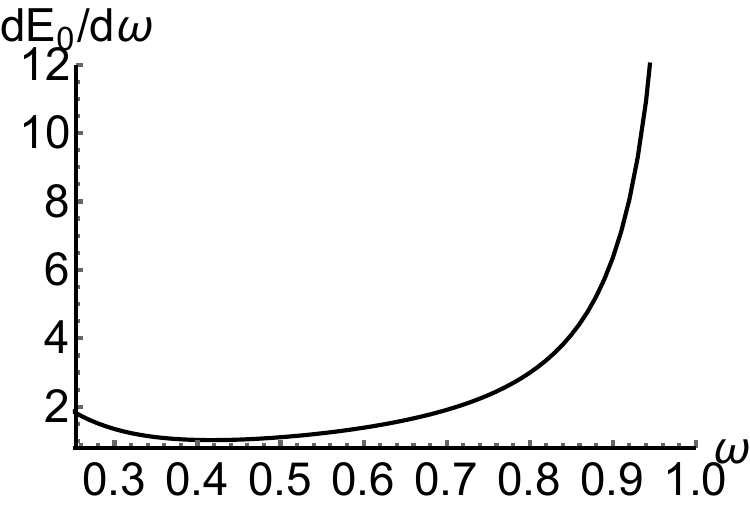}
\caption{Top panel: $\omega$ vs $E_0$ when $\sigma=-1$. Bottom panel: $dE_0/ d\omega$ vs $\omega$ when $\sigma=-1$. }
\label{fig5}
\end{center}
\end{figure}

In the second case ($\sigma=-1$),  we see that solutions are always linearly stable since $dE_0/d\omega>0$ but the variation of pulse power with its frequency shows sudden growth around $\omega=0.98$(Fig. \ref{fig5}). This change actually associated  sudden drop of Fisher information (Fig.\ref{fig3}). As we mentioned, the solution changes its characteristics from regular to flat top bright solitons.  It is worth mentioning that both the solutions are linear stable according to Vakhitov-Kolokolov criterion.

\section{Conclusions}
Information  theory plays important role in the different branches of science because of it usefulness to extract information from  complicated systems. Here  we consider Fisher information (FI) to extract information on the change of characteristics of optical pulse in cubic-quintic media because FI can detect local changes of a distribution. We know that cubic-quintic nonlinear media can hold sharp top (regular) and flat top solitons depending on sign of quintic nonlinearity and frequency of the pulse. In case of focusing we will always have sharp top soliton. This  soliton in cubic-quintic media possesses a definite Fisher information both in coordinate and momentum spaces  even if the soliton enter in the linearly unstable region. In the case of defocusing quintic nonlinearity we can have both regular and flat top soliton. The transition can occurs when the frequency approaches to $1$.  During transition from RS to FTS  we find sudden decrease in the FI and, at the same time  the power starts to grow very fast with frequency. However, it remains linearly stable according to Vakhitov-Kolokolov criterion.

\section*{Acknowledgment}
One of the authors (GAS) would like to acknowledge the funding from the “Science  and Engineering Research Board(SERB), Govt.of India” through Grant No. CRG/2019/000737. Mr. B. Layek would like to thank the Department of Physics, Kazi Nazrul University, India for providing research facility during this work.
\vskip 0.25cm
{\bf Declaration of conflict of interest:}\\
We declare that there is ‘no conflict of interest’ in the work.
\vskip 0.25cm
{\bf Data availability statement:}\\
Data generated or analyzed during this study are provided in full within the article.


\begin{thebibliography}{99}
\section*{References}
\bibitem{rr1} Michael A. Nielsen and Isaac L. Chuang, {Quantum Computation and Quantum Information}, Cambridge University Press, Ed. 10, (2010).
\bibitem{rr2} C. E. Shannon, A mathematical theory of communication, The Bell System Technical Journal, {\bf  27}, 379-423(1948), doi: 10.1002/j.1538-7305.1948.tb01338.x.
\bibitem{rr3} I. Bialynicki-Birula, J. Myceilski, Uncertainty relations
for information entropy in wave mechanics, Commun. Math. Phys. {\bf 44}, 129(1975).
\bibitem{rr4} R. Fisher, Theory of statistical estimation, Mathematical
Proceedings of the Cambridge Philosophical Society {\bf 22}, 700(1925).
\bibitem{rr5} S. Chatterjee and Golam Ali Sekh and B. Talukdar, Fisher Information for the Morse Oscillator, Reports on Mathematical Physics {\bf 85}, 281(2020).
\bibitem{rr6} A. Saha, B. Talukdar and S. Chatterjee, On the realization of quantum Fisher information, Eur. J. Phys. {\bf 38} 025103(2017).
\bibitem{rr7} M. J. W. Hall, Quantum properties of classical Fisher information, Phys. Rev. A {\bf 62} 012107(2000).
\bibitem{rr8} Entropic uncertainty relation and their applications, P. J. Coles, M. Berta, M. Tomamichel and S. Wehner, Rev. Mod. Phys. {\bf 89}, 015002(2017).
\bibitem{rr9} V. Aguiar and I. Guedes, Shannon entropy, Fisher information and uncertainty relation for log-periodic oscillator,
Physica A \textbf{423}, 72(2015).
\bibitem{rr10} V. Aguiar and I. Guedes, Fisher information of
quantum damped harmonic oscillator, Phys. Scr. \textbf{90},
045207(2015).
\bibitem{rr11} V. H. L. Bessa and I. Guedes, The quantum Lane-
Emden-type Kanai-Caldirola oscillators, J. Math. Phys.
\textbf{53}, 122104(2012).
\bibitem{rr12} P. Caldirola, Quantum analysis of modified Caldirola-
Kanai oscillator model for electromagnetic fieldsin timevarying.
IL Nuovo Cimento \textbf{18}, 393(1941).
\bibitem{rr13} A. Plastino, G. Bellomo and A. R. Plastino, On a conjecture regarding Fisher information, Adv. Math. Phys.
\textbf{2015}, 120698(2015).
\bibitem{rr14}P. Sanchez-Moreno, A. R. Plastino and J. S. Dehesa, A
quantum uncertainty relation based on Fisher’s information,
J. Phys. A: Math. Theor. \textbf{44}, 065301(2011).
\bibitem{rr15}A. J. Stam, Some inequalities satisfied by the quantities
of information of Fisher annd Shannon, Inf. Control. \textbf{2},
101(1959).
\bibitem{rr16} W. A. Yahya, K. J. Oyewumi, K. D. Sen, Quantum information entropies for the l-state Poschl–Teller-type potential, J. Math. Chem. {\bf 54}, 1810(2016).
\bibitem{rr16a} B. Layek, S. Sultana and G. A. Sekh, Information measures on matter-wave solitons in Bose-Einstein condensates,  Can. J. of Phys. (2023), https://doi.org/10.1139/cjp-2022-0221.
\bibitem{rr17}  L.J. Song, J. Ma, D. Yan and X.G. Wang, Quantum Fisher information and chaos in the Dicke model, Eur. Phys. J. D {\bf 66}, 201(2012).
\bibitem{rr18} A. Nagy and E. Romera, Fisher information, R\'enyi entropy power and quantum phase transition in the Dicke model, Physica A {\bf 391}, 3650(2012).
\bibitem{rr19} Kajal Krishna Dey, Sudipta Das and Golam Ali Sekh, On the information entropy of matter-waves in quasi-periodic lattice potentials, Eur.  Phys. J. D \textbf{73}, 18(2019).
\bibitem{rr20} Golam Ali Sekh, Aparna Saha, Benoy Talukdar, Shannon entropies and Fisher information of K-shell electrons of neutral atoms, Physics Letters A {\bf 382}  315-320(2018).
\bibitem{rr21} Golam Ali Sekh, B. Talukdar, S. Chatterjee, and B. A. Khan, Information theoretic approach to effects of spin-orbit coupling in quasi-one-dimensional Bose-Einstein condensates,  Phys. Scr. \textbf{97}, 115404(2023).
\bibitem{rr22} N. G. Vakhitov and A. A. Kolokolov, Stationary solutions of the wave equation in the medium with nonlinearity saturation, Radiophys. Quantum Electron.{\bf 16}, 783(1973).
\bibitem{rr23a} G. P. Agrawal, Nonlinear Fiber Optics. 5th. Academic Press, Elsevier, 2012. ISBN: 9780123970237.
\bibitem{rr23} Sudipta Das, Golam Ali Sekh, Dynamics of Compressed Optical Pulse in Cubicquintic Media,Fiber and Integrated Optics, {\bf 39}, 122-136(2020), Doi: 10.1080/01468030.2020.1800143.
\bibitem{rr24} K. H. I. Pushkarov, D.I. Pushkarov, I. V. Tomov, Self-action of light beams in nonlinear media: soliton solutions, Optical and Quantum Electronics {\bf 11} 471(1979).
\bibitem{rr25} M. A. Grado Caffaro and M. Grado Caffaro, Some remarks on qasi-solitons in optical fibers, Active and Passive Elec. Comp., {\bf 15}, 103(1993).
\end{thebibliography}
\end{document}